\documentclass[conference]{IEEEtran}

\usepackage{amsmath}
\usepackage{amssymb}
\usepackage[pdftex]{graphicx}
\usepackage{multicol}
\usepackage{multirow}

\usepackage{epstopdf}
\usepackage{tabularx}
\usepackage{romannum}
\usepackage{gensymb}
\usepackage{caption}
\usepackage{subcaption}
\usepackage{cleveref}

\begin{document}

\title{Investigation of Torque Ripple in Voltage Source Inverter driven Induction Motor Drive operated with Space Vector based Harmonic Elimination Pulse Width Modulation Scheme}

\author{\IEEEauthorblockN{ Rohan Sandeep Burye}
\IEEEauthorblockA{\textit{School of Electrical Sciences} \\
\textit{Indian Institute of Technology Goa,}\\
Ponda, India \\
rohan19242205@iitgoa.ac.in}
\and
\IEEEauthorblockN{Ravi Teja Arumalla}
\IEEEauthorblockA{\textit{Dept. of Electrical and Electronics Engineering} \\
\textit{Rajiv Gandhi University of}
 \\ \textit{Knowledge and Technology,}\\
Ongole, India \\
ravitejaeee.89@gmail.com}
\and
\IEEEauthorblockN{Sheron Figarado}
\IEEEauthorblockA{\textit{School of Electrical Sciences} \\
\textit{Indian Institute of Technology Goa,}\\
Ponda, India  \\
sheron@iitgoa.ac.in}
}
\maketitle

\begin{abstract}
The lower order harmonic elimination using space vector pulse width modulation (SVPWM) technique is possible by introducing the dwell time division coefficient  `$k$' in space vector. In this paper, torque ripple analysis is carried out for space vector based harmonic elimination (SVHE) scheme and the obtained results are compared with the existing schemes such as conventional and advanced bus clamping SVPWM schemes. Further, the obtained results are validated through simulation results. The SVHE scheme gives the lower torque ripple than the conventional and advance bus clamping schemes at higher modulation indices. 

\end{abstract}

\IEEEpeerreviewmaketitle

\section{Introduction}
The torque ripple is a significant performance metric in variable-speed induction motor (IM) drive systems\cite{Hari2015}. According to \cite{Tripathi2015}, the torque ripple in an IM drive is caused by the harmonic content in an inverter's output waveform. In \cite{Narayanan2005}, a time-domain method of analysis based on the concept of stator flux ripple is proposed to analyse the harmonic distortion caused by space vector based PWM approaches. In addition, \cite{Tripathi2019} proposes an analytical method for predicting torque ripple of an IM drive running at low pulse number. The selection of pulse width modulation (PWM) techniques to trigger the inverter switches determines the quality of the inverter's output waveform \cite{Arumalla2020}. Over these years various PWM schemes are devised for low, medium and high  frequency applications. An extensive review on various PWM techniques for motor drive application is presented in \cite{Henglin2016}.

One of the most popular PWM technique is selective harmonic elimination (SHE). The SHE permits elimination of certain lower order harmonics which are dominant component of the torque ripple. It is especially useful in high power, low frequency application \cite{Dahidah2015}.  The SHE requires solving the transcendental equations obtained through fourier analysis. Hence, the real time implementation of SHE is challenging task in dynamic conditions. Using Groebner bases and symmetric polynomial theory, an algebraic method is proposed to solve transcendental equations associated with SHE in \cite{Yang2016}. Unlike traditional numerical methods, this method gives all the possible solutions and had no requirements on choosing initial value.  

The most widely used PWM method for real time implementation is sinusoidal pulse width modulation (SPWM). In this method a sinusoidal reference waveform is compared with a high frequency carrier waveform to generate PWM. The advantage of this method is its simplicity and ease of implementation with microcontroller. By adding appropriate common mode voltage it is possible to generate a conventional and bus clamping SVPWM sequences. A phase shifted SPWM to reduce common mode voltage in multi-phase two level inverter is proposed in \cite{Liu2016}. A FPGA-based implementation of SPWM strategy for high frequency inverter is presented in \cite{Lakka2014}. A zero voltage switching based SPWM modulation scheme is proposed for grid connected Inverter in \cite{Chen2016}. The analytical and graphical method to generate carrier-based PWM is extensively reviewed in \cite{Hava1999}.


The space vector pulse width modulation(SV-PWM) is an immensely popular method for a three phase 2- level VSI.  Both carrier based and space vector based PWM can implement conventional and clamping sequences. But SPWM fails to implement advance bus clamping (ABC-PWM) sequences. Therefore, space vector based PWM is more general than triangle comparison method of generating PWM. Moreover, SVPWM has higher DC bus utilization than SPWM. The SVPWM techniques depend on the sequence in which the space vectors are applied. An overview of different SVPWM techniques can be found in \cite{Nandhini_2020}. The most popular space vector based PWM techniques are conventional space vector PWM(CSV-PWM), bus clamping PWM(BC-PWM) and advance bus clamping PWM(ABC-PWM).

In the CSV-PWM, the dwell time associated with the zero vector is divided equally among zero vectors $V_0$ and $V_7$ . The CSV-PWM has 3 switchings per subcycle. The CSV-PWM offers the best total harmonic distortion(THD) performance. The bus clamping sequences utilize two active vector and a zero vector in a sector resulting into 2 switchings per subcycle. The bus clamping sequences clamp one of the phase to the DC bus for a 60\degree duration in a subcycle which leads to lower switching losses and reduction in current ripple of an inverter \cite{Kumar2018}. With respect to the clamping duration, bus clamping is further classified as continual and split clamping. An analysis based on the torque ripple and harmonic distortion of continual clamp and split-clamp PWM for IM drive is shown in \cite{Das2019}. It is shown that optimal positioning of the clamping interval in split-clamp PWM yield better results in terms of lower  torque ripple and THD in line current than all other bus clamping sequences.  In Advance bus clamping sequences an active vector is applied twice in a subcycle and it has shown to reduce THD compared to the conventional and bus clamping switching sequence at higher modulation indices \cite{Bhavsar2009}.
 
 The ABC-PWM can offer improved torque ripple profile at higher modulation index. Thus, the space vector based hybrid PWM technique is proposed in \cite{Hari2015}, wherein combination of CSV-PWM and ABC-PWM sequences is used for different regions of a sector in the space vector diagram. In \cite{Basu2009}, a PWM technique is proposed that is a combination of optimal continuous modulation and discontinuous modulation and has been demonstrated to reduce torque ripple by 30\%. In \cite{Zhao2010}, four space-vector-based hybrid PWM approaches are suggested, which reduce line current distortion and switching loss in IM drives when compared to CSV-PWM.

The unequal division of the dwell times associated with the space vectors in SVPWM technique can lead to sequences having less torque ripple for certain modulation indices. Moreover, this dwell time division also makes elimination of particular lower order harmonics possible which result in lower torque ripple. The Space Vector sequences generated using this concept of dwell time arrangement are called space vector harmonic elimination sequences (SVHE). The SVHE offers advantage of both SHE and SVPWM techniques. The concept of eliminating harmonics using space vector first appeared in \cite{Grewal2000}. A model predictive switching pattern control to eliminate the lower order harmonics based on space vector approach was proposed in \cite{Gao2017} for current source converter.
 
The objective of this paper is to perform the torque ripple analysis on a SVHE sequence which is formulated by using the concept of dwell time rearrangement to selectively eliminate a particular lower order harmonic\cite{Arumalla2021}. The performance of this sequence with regards to torque ripple are compared with the CSV-PWM and ABC-PWM techniques. All of the switching sequences under consideration are having 18 switchings/phase/fundamental cycle which ensures near equal switching loss and makes the comparison fair. The analytically obtained and simulated torque ripple waveforms are presented. 

\begin{figure}[ht]
\centering
\includegraphics[scale=0.35]{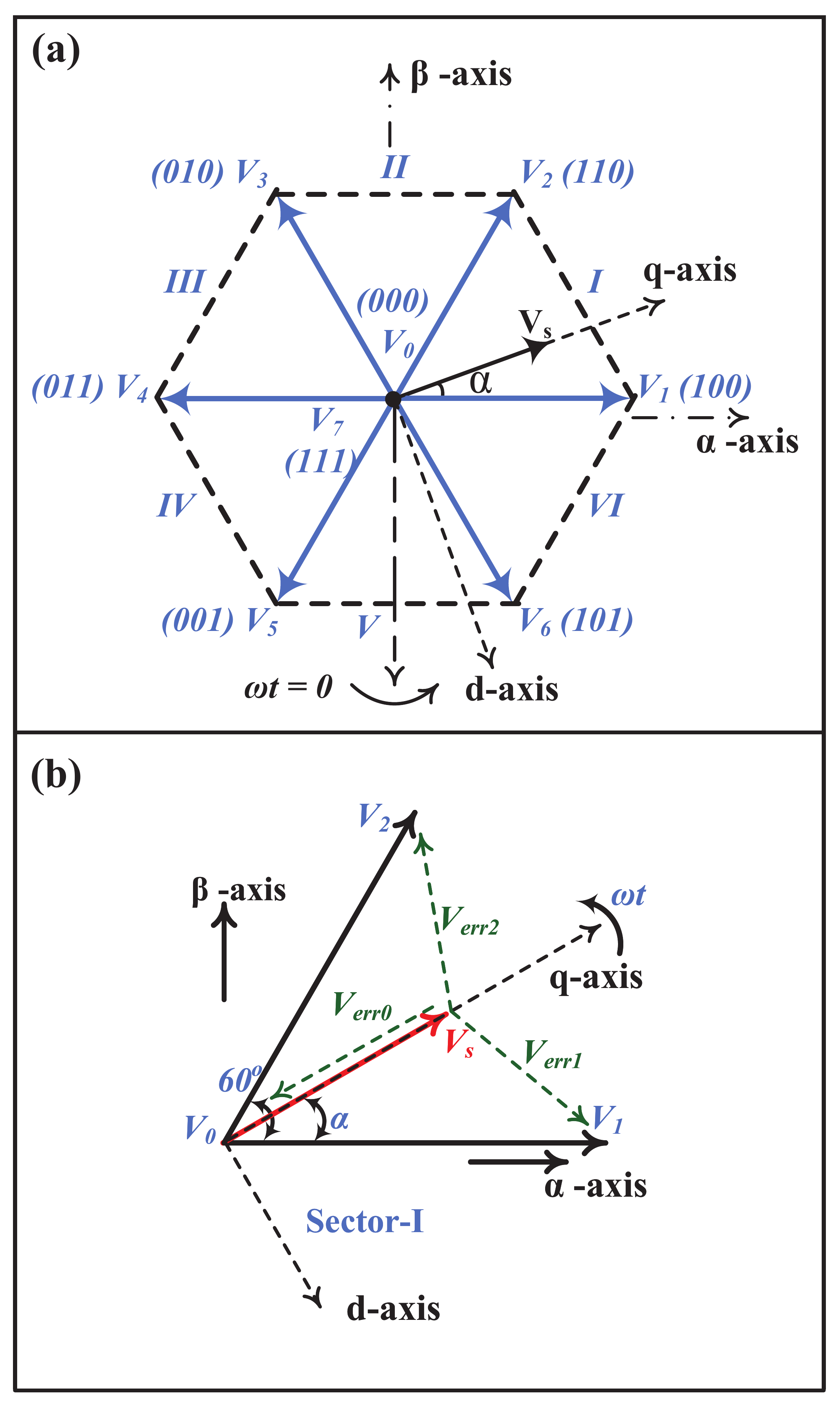}
\caption{(a)SV Diagram (b)Error voltage vectors in sector-1}
\label{sv}
\end{figure}

\section{Analysis of Torque Ripple}
The Fig. \ref{sv}(a) shows  Space Vector diagram for a three phase 2 level VSI. It consist of six active vectors ($V_1\mbox{-}V_7$) and zero vectors ($V_0$ and $V_7$) forming a Hexagon. The  Sector-1 of SV diagram along with the error voltages are shown in Fig. \ref{sv}(b). Any reference vector($V_{ref}$) within the sector-1 is realized by switching active vectors $V_1,V_2$ and zero vectors $V_0,V_7$ for duration $T_1,T_2$ and $T_0$ respectively. This is called volt-sec balance. The dwell timings are calculated as

\begin{subequations}
\begin{align}
&T_1=\frac{V_{ref}sin(60\degree-\alpha)}{V_{dc}sin(60\degree)}T_s \\
&T_2=\frac{V_{ref}sin(\alpha)}{V_{dc}sin(60\degree)}T_s \\
&T_0=T_s-T_1-T_2 \\
&T_s=\frac{1}{300nm} \\
&V_{ref}=0.866mV_{dc}
\end{align}
\end{subequations}

\begin{align*}
\text{where,}\\
&m=\text{modulation index} \\
&n=\text{number of sample in a sector} \\
&T_s= \text{sample duration} \\
&V_{dc}= \text{DC link voltage of the Inverter}
\end{align*}

The different SVPWM sequences under consideration are tabulated in  Table \ref{tb1}. 
\begin{table}
\caption{SVPWM Sequences under study}
\label{tb1}
\begin{tabular}{c c c c} \hline
Sequence & SPS & Sequence & Dwell Time \\ 

 &  &  & Arrangement \\ 
\hline
CSV & 3 & 0127-7210- & $T_0.T_1.T_2.T_7\mbox{-}$  \\
&$(10\degree,30\degree,50\degree)$ &0127 & $T_7.T_2.T_1.T_0\mbox{-} $  \\
 & & & $T_0.T_1.T_2.T_7 $ \\ \hline

 ABC-\Romannum{1} & 2 & 0121-1210 & $T_0.0.5T_1.T_2.0.5T_1\mbox{-}$  \\
  &$(15\degree,45\degree)$ & & $0.5T_1.T_2.T_1.0.5T_0 $ \\ \hline
   ABC-\Romannum{2} & 2 & 7212-2127 & $T_7.0.5T_2.T_1.0.5T_2\mbox{-}$ \\
   &$(15\degree,45\degree)$ & & $T_2.T_1.0.5T_2.T_7 $  \\ \hline
SVHE & 2 & 0121-7212 & $T_0.kT_1.T_2.(1-k)T_1\mbox{-}$  \\ 
& $(15\degree,45\degree)$ & & $T_7.(1-k)T_2.T_1.kT_2 $  \\ \hline
\end{tabular}
\end{table}

In CSV, the zero vector dwell time is divided equally among $T_0$ and $T_7$ which gives best THD results. The ABC-\Romannum{1}  and ABC-\Romannum{2} \ are double switching sequences in which active vector is applied twice in a subcycle. The SVHE sequence proposed in \cite{Arumalla2021} uses a dwell time re-arrangement with a factor $k$ to selectively eliminate either a $5^{th}$ or $7^{th}$ harmonic. The value of $k$ is found by solving the non-linear algebraic equations obtained from the fourier analysis of the pole voltage \cite{Arumalla2021}. The sample per sector(SPS) and the dwell time arrangement for each of the sequences is also shown in Table \ref{tb1}.

The torque ripple waveforms are obtained using the notion of stator flux ripple \cite{Narayanan2005}. In this method, the error voltage vector is resolved  along q-axis and the d-axis of a synchronously rotating reference frame, with the q-axis aligned along  $V_{ref}$ (see Fig. \ref{sv}). The q-axis component of error voltage vector for SVHE sequence is derived and given by \cref{err,err2}. Similarly, it can be obtained for other sequences listed in Table \ref{tb1}.

\begin{equation}\label{err}
  \widetilde{v}_{q,0121}=
    \begin{cases}
      K_1 & 0<t<T_0\\
      K_2+K_1 & T_0<t<T_0+kT_1\\
      K_3+K_1 & T_0+kT_1<t<T_0+kT_1+T_2 \\
      K_2+K_1 & T_0+kT_1+T_2<t<T_s
    \end{cases}       
\end{equation}

\begin{equation}\label{err2}
  \widetilde{v}_{q,7212}=
    \begin{cases}
      K_1& T_s<t<T_s+T_7\\
      K_3+K_1  & T_s+T_7<t<T_s+T_7+(1\mbox{-}k)T_2\\
        K_2+K_1  & T_s+T_7+(1\mbox{-}k)T_2<t<2T_s\mbox{-}kT_2 \\
  K_3+K_1 & 2T_s-kT_2 <t<2T_s
    \end{cases}       
\end{equation}

\begin{align*}
\text{where,}\\
& K_1=-V_{ref}\\
&K_2=V_{dc}cos(\alpha) \\
&K_3=V_{dc}cos(60\degree-\alpha) 
\end{align*}

The time integral of $\widetilde{v}_q$ gives the q-axis stator flux ripple ($\widetilde{\psi}_q$), which is an indicator of q-axis current ripple $\widetilde{i}_q$. The analytical expression for q-axis stator flux ripple for the SVHE is evaluated and obtained in \cref{flux,flux2}. 

\begin{equation}\label{flux}
  \widetilde{\psi}_{q,0121}=
    \begin{cases}
      K_1t & 0<t<T_0\\

      K_4t \mbox{-} K_2T_0  & T_0<t<T_0+kT_1\\
      \begin{cases}
      K_2(kT_1)+(K_5)t \\ \mbox{-}K_3(T_0+kT_1) 
         \end{cases} & 
          \begin{cases}
          T_0+kT_1<t \\ <T_s\mbox{-}(1\mbox{-}k)T_1
           \end{cases}\\
      \begin{cases}
      (K_4)(t\mbox{-}T_2) \\ \mbox{-}K_2T_0+(K_5)T_2 
      \end{cases} & T_s\mbox{-}(1\mbox{-}k)T_1<t<T_s
    \end{cases}       
\end{equation}

\begin{figure*}[ht]
\centering
\includegraphics[scale=0.50]{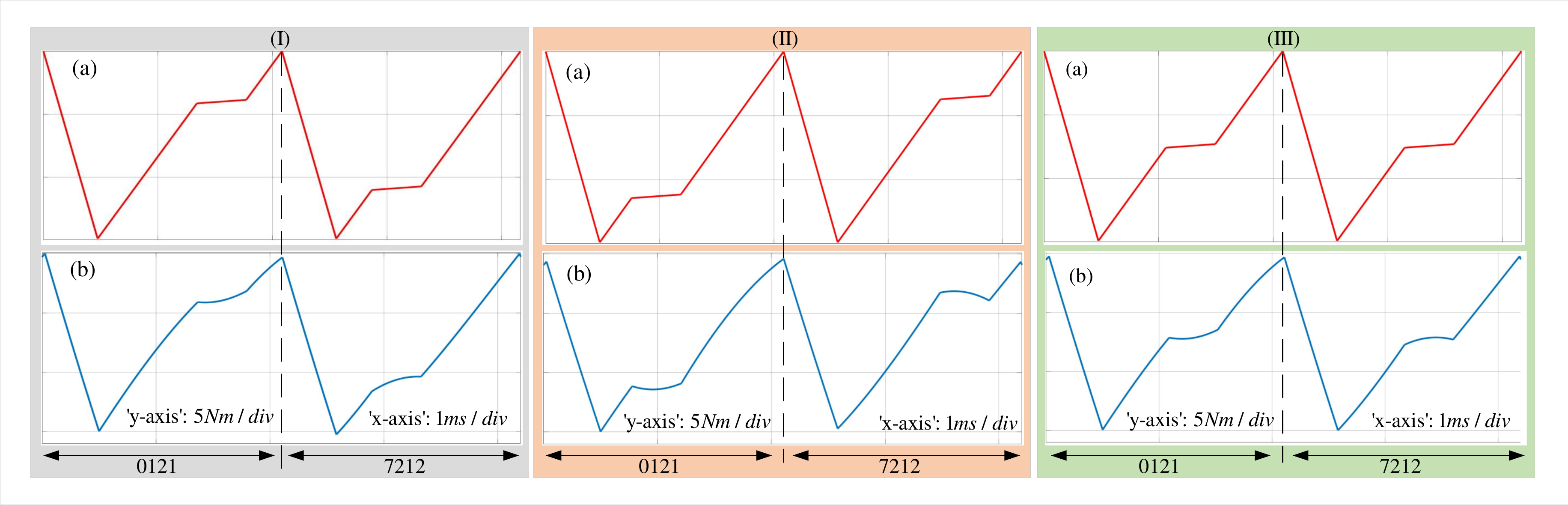}
\caption{ Torque ripple waveforms of SVHE sequence 0121-7212: (a)analytical  \& (b)simulation  for (\Romannum{1}) $5^{th}$ elimination, (\Romannum{2}) $7^{th}$ elimination \& (\Romannum{3}) No harmonic elimination ($k=0.5$) case at 0.8 modulation index}
\label{tr0}
\end{figure*}

\begin{equation}\label{flux2}
  \widetilde{\psi}_{q,7212}=
    \begin{cases}
      K_1(t\mbox{-}T_s) & T_s<t<T_s+T_7\\

      (K_5)(t\mbox{-}T_s)\mbox{-}K_3T_7   &
       \begin{cases}
       T_s+T_7<t<T_s \\ +T_7+(1\mbox{-}k)T_2
       \end{cases}\\
       
      \begin{cases}
      (K_6)(T_7+(1\mbox{-}k)T_2) \\ \mbox{-}K_3T_7+(K_4)(t\mbox{-}T_s)  
         \end{cases} & 
          \begin{cases}
          T_s+T_7+(1\mbox{-}k)T_2 \\ <t<2T_s\mbox{-}kT_2
          \end{cases} \\
      \begin{cases}
      (K_5)(t\mbox{-}T_s\mbox{-}T_1) \\ \mbox{-}K_3T_7+(K_4)T_1  
      \end{cases} & 2T_s-kT_2 <t<2T_s
    \end{cases}       
\end{equation}

\begin{align*}
\text{where,}\\
&K_4=K_2+K_1 \\
&K_5=K_3+K_1 \\
&K_6=K_3-K_2
\end{align*}

Similarly, the stator flux ripple for other sequences listed in Table \ref{tb1} is obtained. Torque  rippple is produced when $i q$ interacts with fundamental flux along the d-axis, according to \cite{Hari2015}. As a result, the ripple in an Induction motor's produced torque ($\widetilde{\tau}_d$) is given by 
\begin{equation}\label{Td}
\widetilde{\tau}_d=\frac{2}{3}\cdot\frac{P}{2}\cdot\frac{V_{ref}}{2\pi f_1}\cdot \frac{1}{L_O}\cdot \left[\frac{1}{\sigma_s+\sigma_r}-1 \right]\widetilde{\psi}_q
\end{equation}
Here,$\sigma$ is the ratio of total leakage inductance to the  magnetizing inductance of the motor\cite{Basu2009}. 

The parameters of the Induction motor used for the simulation are given in Table \ref{tb2}.

\begin{table}
\caption{Parameters of Induction Motor} \label{tb2}
\begin{center}
\begin{tabular}{|c|c|} \hline
\textbf{Parameter} & \textbf{Value}  \\ \hline
Rated Power, $P_{rated}$ & 7.5 $kW$  \\ \hline
Stator Resistance, $R_s$ & 1.1667 $\ohm$  \\ \hline
Rotor Resistance, $R_r$ & 3.2105 $\ohm$ \\ \hline
Magnetizing Inductance , $L_O$ & 0.3025 $H$ \\ \hline
 Stator leakage coeffcient, $\sigma_s$ & 0.0392 \\ \hline
 Rotor leakage coeffcient, $\sigma_r$ & 0.0392 \\ \hline
\end{tabular}
\end{center}
\end{table}

By inspection of torque ripple waveform for SVHE and CSV sequence, it is inffered that the torque ripple for SVHE depends solely on the $T_0$ whereas in case of CSV it depends on the $T_0$ and $T_2$. This is true for the uniform sampling locations as specified in Table \ref{tb1}. Thus, an expression for peak to peak q-axis stator flux ripple is derived for SVHE and given by (\ref{dpsi1}).
\begin{equation} \label{dpsi1}
\begin{split}
\Delta\widetilde{\psi}_{q,SVHE}&= V_{ref}\cdot T_{0,SVHE}\\
 & = V_{ref}T_{s1}(1-m(sin(\alpha_1)+sin(60\degree-\alpha_1)))
\end{split}
\end{equation}

\begin{align}
&\text{where,} \nonumber\\
&\text{Sample-1 duration of SVHE}(T_{s1})=\frac{1}{600m} \nonumber \\
&\text{Sample-1 location of SVHE}(\alpha_1)=15\degree \nonumber
\end{align}

Similarly, expression for the peak to peak q-axis stator flux ripple is derived for CSV and given by (\ref{dpsi2}).
\begin{equation} \label{dpsi2}
\begin{split}
\Delta\widetilde{\psi}_{q,CSV}&= V_{ref}\cdot T_{0,CSV}+2(V_{ref}-V_{dc})T_{2,CSV}\\
 & = T_{s2}\times[V_{ref}(1-m(sin(\alpha_2)+sin(60\degree-\alpha_2))\\
& \qquad+(V_{ref}-V_{dc}cos(60\degree-\alpha_2))2msin(\alpha_2)]
\end{split}
\end{equation}

\begin{align}
&\text{where,} \nonumber\\
&\text{Sample-1 duration of CSV}(T_{s2})=\frac{1}{900m} \nonumber \\
&\text{Sample-1 location of CSV}(\alpha_2)=10\degree \nonumber
\end{align}
The magnitude of peak-to-peak torque developed($\Delta\widetilde{\tau}_d$) corresponding to these stator flux ripple is obtained from (\ref{Td}). The  expression for range of modulation indices for which $\Delta\widetilde{\psi}_{q,SVHE}<\Delta\widetilde{\psi}_{q,CSV}$ and is obtained in (\ref{m}).

\begin{equation}\label{m}
m>\dfrac{0.5+\tfrac{2sin(\alpha_2)cos(60\degree-\alpha_2)}{cos(30\degree)}}{1.5(sin(\alpha_1)+sin(60\degree\mbox{-}\alpha_1))+sin(\alpha_2)-sin(60\degree\mbox{-}\alpha_2)}
\end{equation}

The torque ripple in IM drive is independent of the d-axis ripple and is related to the q-axis ripple.
However, $F_d$ and $F_q$, where $F_d$ and $F_q$ are the RMS of the d-axis ripple and q-axis flux ripple waveforms, respectively, have an equivalent impact on the THD of the motor current waveform. Thus, a decrease in q-axis ripple denotes a decrease in both THD and in torque ripple, but a decrease in d-axis ripple solely denotes a decrease in THD.

The THD in the motor current is a useful indicator of current ripples \cite{Narayanan2005}. The current THD is defined as (\ref{thd}), where $I_1$ and $I_n$ are the RMS of the no load current's fundamental and $n^{th}$ harmonic component, respectively.
\begin{equation}\label{thd}
I_{THD}=\frac{1}{I_1} \sum_{n=2}^{\infty} I_n 
\end{equation}
Unlike q-axis ripple, the peak-to-peak magnitude of the d-axis ripple ($\Delta\widetilde{\psi}_{d}$) is independent of the modulation index for SVHE and CSV sequences . The $\Delta\widetilde{\psi}_{d}$ for SVHE is found to be lower than the CSV sequence. Although the $\Delta\widetilde{\psi}_{d}$ is independent of the dwell time arrangement, the  $F_d$  depends on the dwell time division coeffcient `$k$'.

\begin{table}
\caption{Analytical results of torque ripple magnitude for Sequence-CSV:0127-7210-0127,ABC-\Romannum{1}:0121-1210,ABC-\Romannum{2}:7212-2127, and SVHE:0121-7212} \label{tb3}
\begin{center}
\begin{tabular}{|c|c|c|c|c|c|c|} \hline
\multirow{3}{*}{Modulation } & \multicolumn{6}{c|}{Peak-to-peak ripple magnitude($Nm$)}  \\ \cline{2-7}  
& CSV   & ABC-\Romannum{1}  & ABC-\Romannum{2}  & \multicolumn{3}{c|}{SVHE}\\ \cline{2-7}
 
Index & $H_{NE}$   & $H_{NE}$   & $H_{NE}$   & $H_{NE}$   & $H_{5}$ & $H_{7}$ \\  \hline

0.6 & 20.066 &  55.046 & 55.07 & 27.57 & 27.57 & 27.57 \\ \hline
0.65 & 17.011 & 48.547 & 48.618 & 24.4 & 24.4 & 24.4 \\ \hline
0.7 & 14.958 & 42.442 & 42.432 & 21.24 & 21.24 & 21.24 \\ \hline
0.75 & 13.025 & 36.066 & 36.038 & 18.06 & 18.06 & 18.06 \\ \hline
0.8 & 11.7 & 29.723 & 29.588 & 14.9 & 14.9 & 14.9 \\ \hline

0.85 & 10.432 & 23.719 & 23.741 & 11.74 & 11.74 & 11.74 \\ \hline

0.9 & 9.136 & 17.763 & 17.812 & 8.569 & 8.569 & 8.569 \\ \hline

0.95 & 7.844 & 11.893 & 11.908 & 5.398 & 5.756 & 5.942 \\ \hline
\multicolumn{7}{l}{$H_{NE}$: No harmonic  elimination ($k=0.5$)} \\
\multicolumn{7}{l}{$H_5$: $5^{th}$ harmonic elimination} \\
\multicolumn{7}{l}{$H_7$: $7^{th}$ harmonic elimination} 
\end{tabular}
\end{center}
\end{table}

\begin{table}
\caption{Simulation results of line current THD at no load for Sequence-CSV:0127-7210-0127 and SVHE:0121-7212} \label{tb4}
\begin{center}
\begin{tabular}{|c|c|c|c|c|} \hline
\multirow{3}{*}{Modulation } & \multicolumn{4}{c|}{Total harmonic distortion(\%)}  \\ \cline{2-5}  
& CSV  & \multicolumn{3}{c|}{SVHE}\\ \cline{2-5}
 
Index & $H_{NE}$  & $H_{NE}$   & $H_{5}$ & $H_{7}$ \\  \hline

0.6 & 71.84  & 87.83 & 84.57 & 97.71 \\ \hline
0.65 & 68.32 & 79.67 & 75.96 & 90.96 \\ \hline
0.7 & 65.12  & 71.85  & 67.58 & 84.34 \\ \hline
0.75 & 62.32  & 64.45 & 59.48 & 78.23 \\ \hline
0.8 & 59.95  & 57.82 & 52.01 & 72.64 \\ \hline

0.85 & 58.06  & 52.22 & 45.36 & 67.68 \\ \hline

0.9 & 56.9  & 47.62 & 39.67 & 63.31 \\ \hline

0.95 & 56.43  & 44.84 & 35.78 & 60.03 \\ \hline
\multicolumn{5}{l}{$H_{NE}$: No harmonic elimination ($k=0.5$)} \\
\multicolumn{5}{l}{$H_5$: $5^{th}$ harmonic elimination} \\
\multicolumn{5}{l}{$H_7$: $7^{th}$ harmonic elimination} 
\end{tabular}
\end{center}
\end{table}

\begin{figure}[h!]
\centering
\includegraphics[scale=0.60]{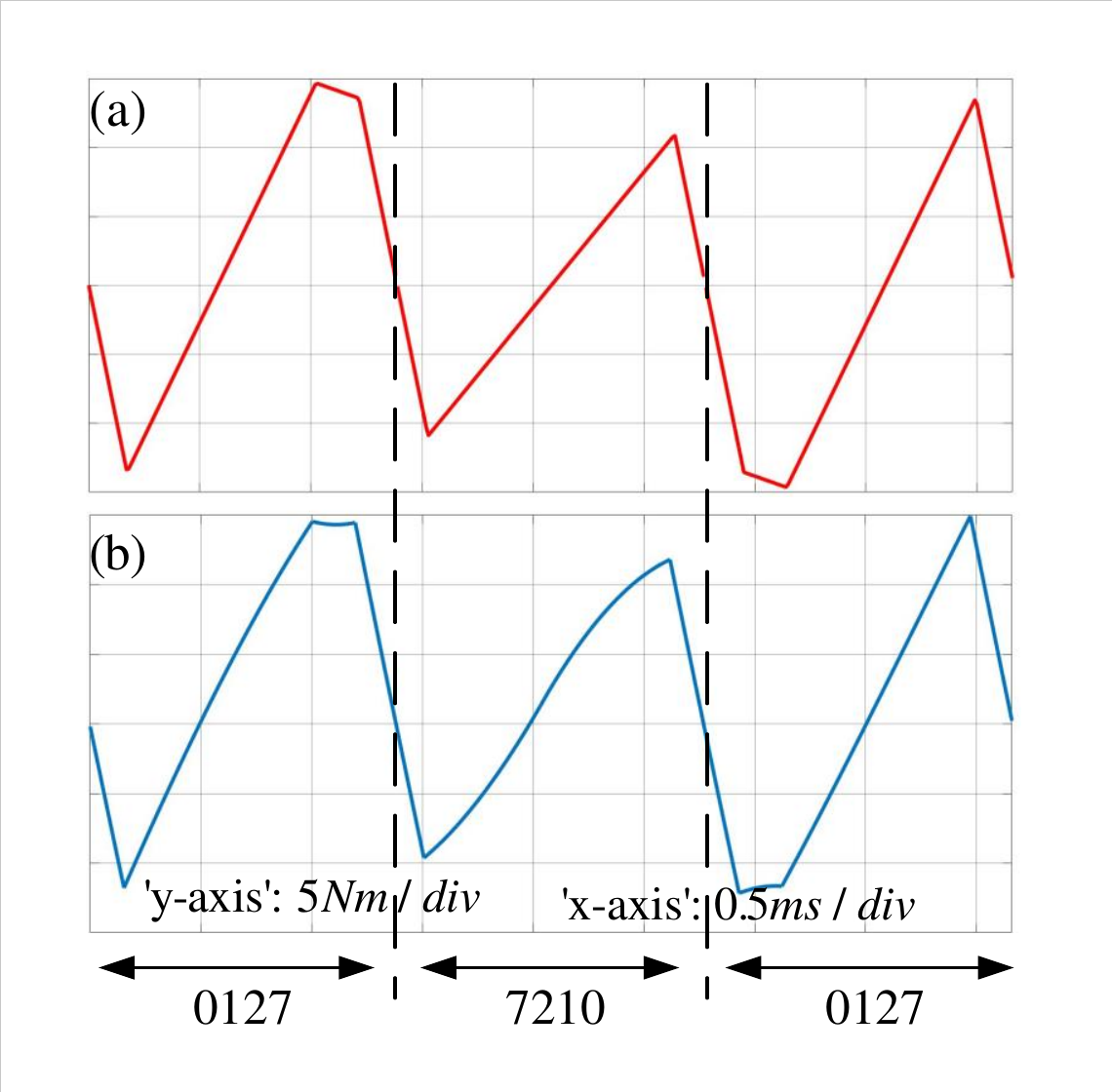}
\caption{ Torque ripple waveforms of CSV sequence 0127-7210-0127: (a)analytical  \& (b)simulation at 0.8 modulation index}
\label{tr1}
\end{figure}

\begin{figure}[h!]
\centering
\includegraphics[scale=0.60]{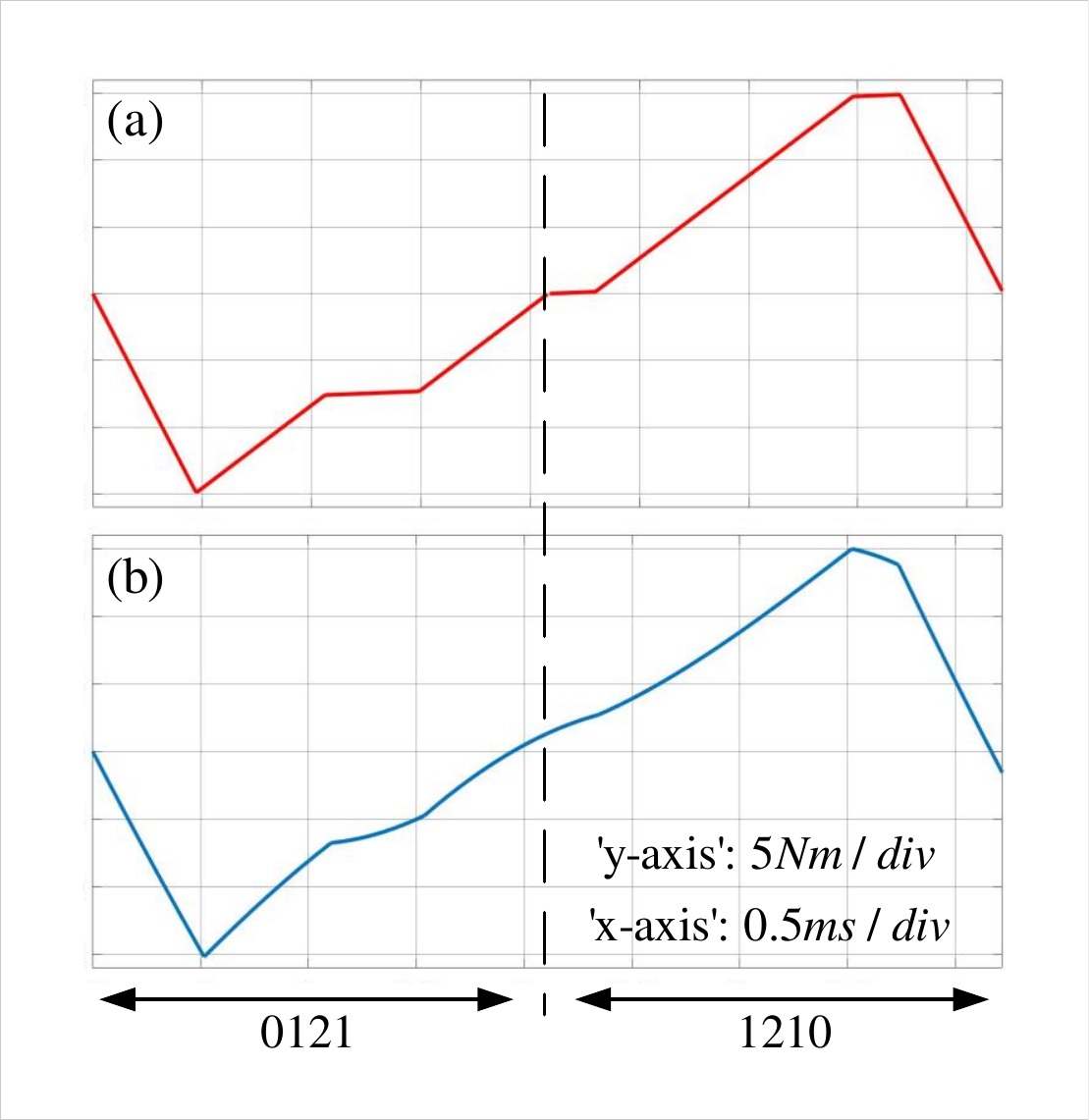}
\caption{ Torque ripple waveforms of ABC-\Romannum{1} sequence 0121-1210: (a)analytical  \& (b)simulation at 0.8 modulation Index}
\label{tr2}
\end{figure}

\begin{figure}[h!]
\centering
\includegraphics[scale=0.60]{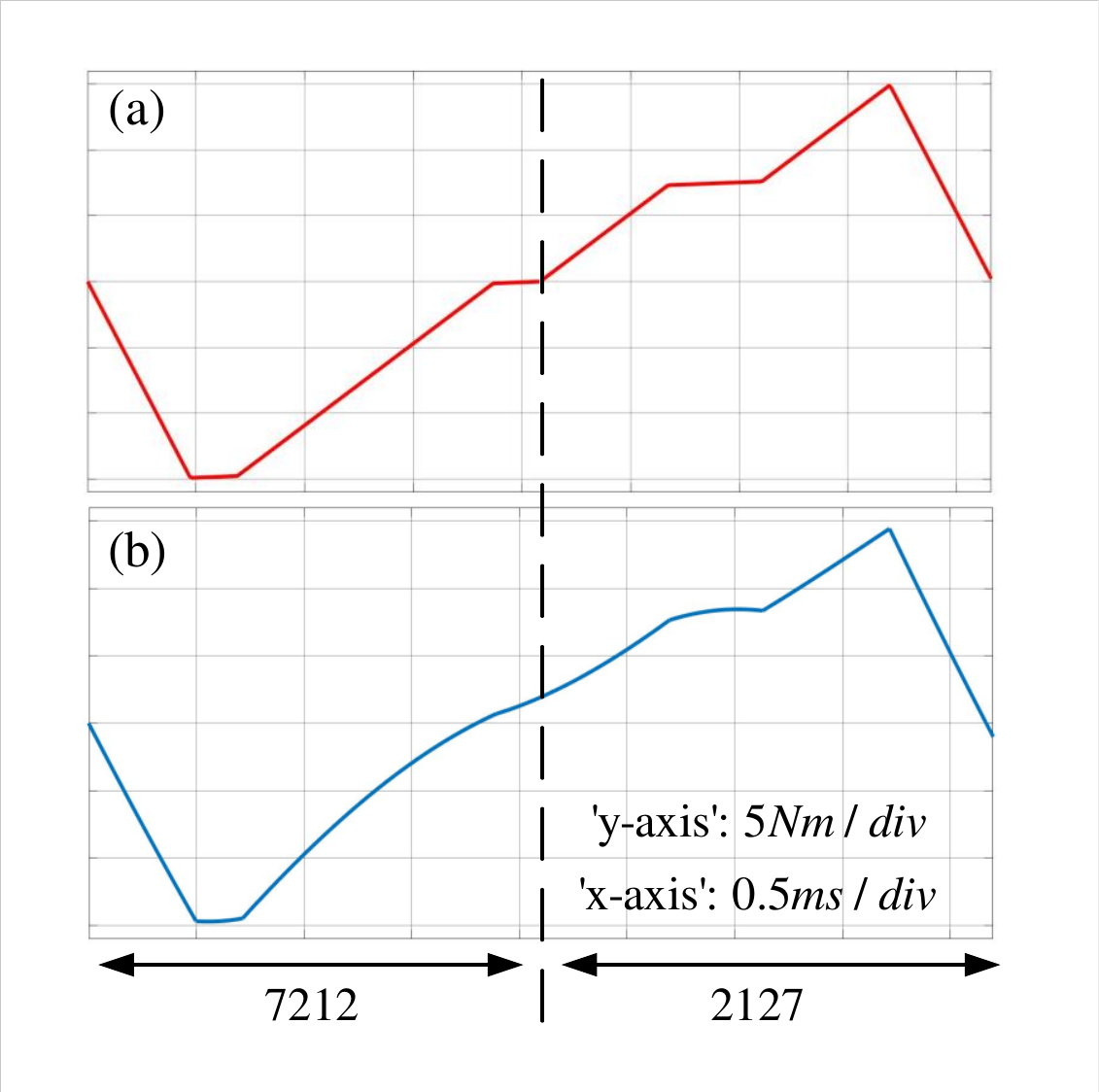}
\caption{ Torque ripple waveforms of ABC-\Romannum{2} sequence 7212-2127: (a)analytical  \& (b)simulation at 0.8 modulation index}
\label{tr3}
\end{figure}

\begin{figure}[h!]
\centering
\includegraphics[scale=0.3]{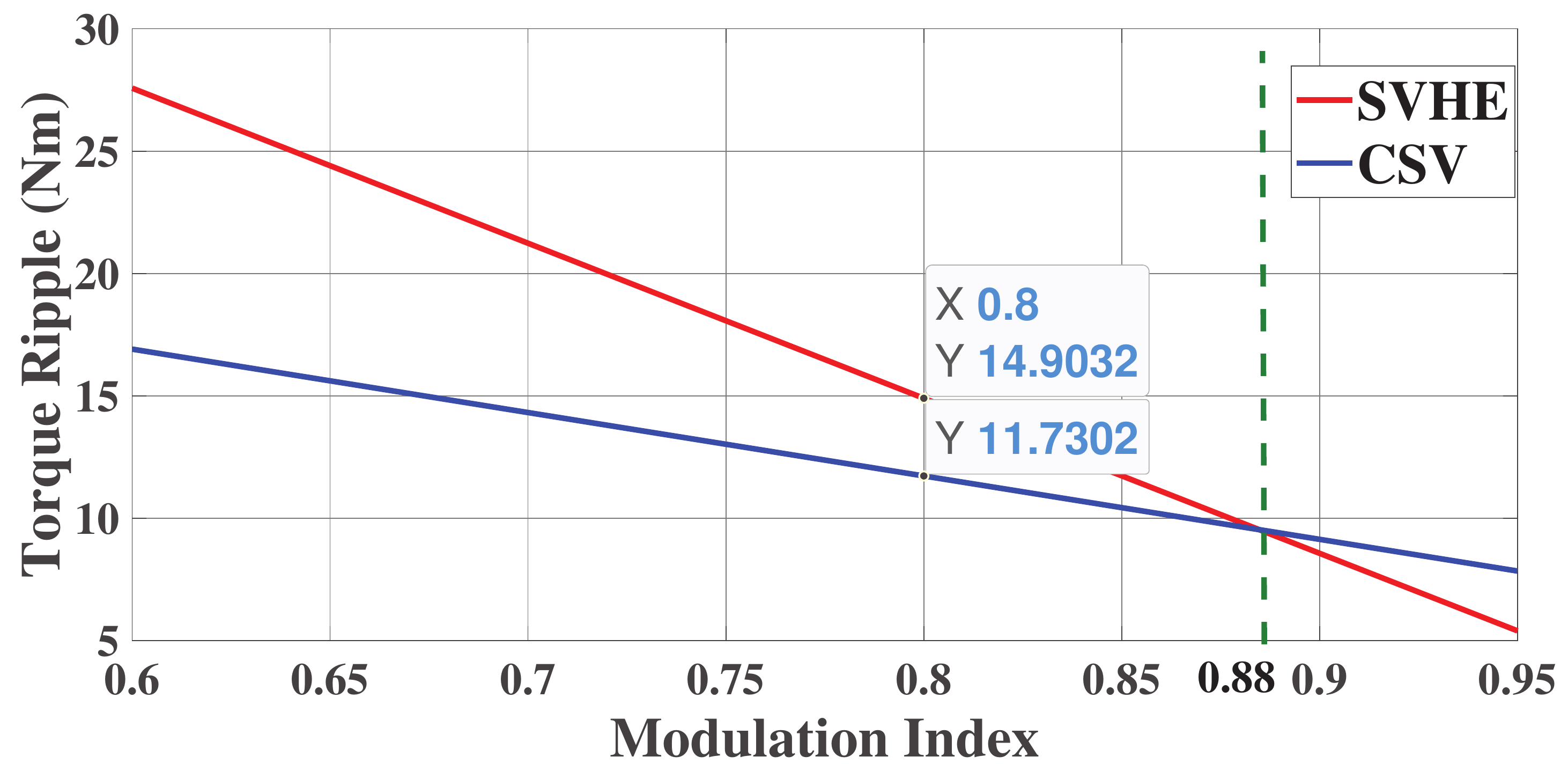}
\caption{ Peak-to-peak torque ripple maginitude variation with modulation index for SVHE and CSV PWM sequences}
\label{tr4}
\end{figure}

\section{Results and Discussion}
The analytical expression for $\widetilde{\psi}_q$ is evaluated for each of the SV-PWM sequences listed in Table \ref{tb1} and corresponding torque ripple waveform $\widetilde{\tau}_d$ is obtained using Symbolic Math in MATLAB script. The analytically obtained peak-to-peak magnitude of torque ripple  is compiled and the result is presented in Table \ref{tb3}.

The torque ripple waveforms are also obtained from the IM model in MATLAB Simulink for these sequences and it is found in good agreement with the analytically obtained waveforms. The Fig. \ref{tr0}-\ref{tr3} shows the torque ripple waveforms obtained from the analytical method and MATLAB simulation for one sector duration for a modulation index of 0.8 and rated power of $7.5kW$. The sector duration is $2T_s$ for 2 SPS(2 sample per sector) and $3T_s$ for 3 SPS. This is valid only for uniform sampling case wherein duration between two consecutive samples is constant. Further the line current THD at no load condition is computed from the simulation, and the results are tabulated in Table \ref{tb4}. The following observations are made from \cref{tb3,tb4}:

\begin{itemize}
\item{The SVHE sequence shows no change in the torque ripple magnitude for elimination and `no elimination' case. This is due to relationship of magnitude with $T_0$ duration alone as shown in (\ref{dpsi1}) and the dwell time re-arrangement happens for the active vector duration.}
\item{The SVHE  gives better torque ripple performance compared to ABC-\Romannum{1} and ABC-\Romannum{2} for entire range of modulation Index.}
\item{SVHE sequence gives better torque ripple than CSV at higher modulation index above 0.8847 as obtained by (\ref{m}) and evident from Fig. \ref{tr4}.}
\item{The torque ripple magnitude decreases with increase in modulation index for all the sequences. This is because as we increase the modulation index, the fundamental magnitude increases relative to harmonics.}
\item{The $I_{THD}$ for $5^{th}$ harmonic elimination case in SVHE shows better results than than CSV for modulation index above 0.7. This is due to the dependency of  $F_d$ on `$k$'}
\end{itemize}

\section{Conclusion}
A torque ripple analysis using the notion of stator flux is presented. The torque ripple for a special Space Vector sequence called SVHE is evaluated and compared with the CSV-PWM and ABC-PWM. This special sequence can eliminate a particular harmonics using the dwell time rearrangement. The results from analytical and simulation are compared and validated. An analytical expression for the modulation index is derived above which the SVHE gives better results than CSV with respect to the torque ripple. The SVHE  has shown better torque ripple performance compared to CSV-PWM at modulation index above 0.8847. The torque ripple are significantly reduced at higher modulation index as evident from the torque ripple waveforms. The SVHE with $5^{th}$ harmonic elimination gives better $I_{THD}$ than CSV for modulation index above 0.7.




\end{document}